\documentclass[epj]{webofc}
\usepackage[utf8]{inputenc}
\usepackage[varg]{txfonts}   
\usepackage{booktabs}
\usepackage{xcolor}
\usepackage{wrapfig}

\definecolor{darkred}{rgb}{0.4,0.0,0.0}
\definecolor{darkgreen}{rgb}{0.0,0.4,0.0}
\definecolor{darkblue}{rgb}{0.0,0.0,0.4}
\usepackage[bookmarks,linktocpage,colorlinks,
    linkcolor = darkred,
    urlcolor  = darkblue,
    citecolor = darkgreen]{hyperref}
\usepackage{subfigure}
\wocname{EPJ Web of Conferences}
\woctitle{Lattice2017}
\begin{document}
%
\selectlanguage{english}
\title{%
Equation of state of non-relativistic matter from automated\\
perturbation theory and complex Langevin
}
\author{%
\firstname{Andrew C.} \lastname{Loheac}\inst{1}\fnsep\thanks{Speaker, \email{loheac@live.unc.edu}} \and
\firstname{Jens} \lastname{Braun}\inst{2} \and
\firstname{Joaqu\'{i}n E.}  \lastname{Drut}\inst{1}\fnsep
}
\institute{%
Department of Physics and Astronomy, University of North Carolina, Chapel Hill, North Carolina 27599, USA
\and
Institut f\"{u}r Kernphysik (Theoriezentrum), Technische Universit\"{a}t Darmstadt, D-64289 Darmstadt, Germany
}
\abstract{%
We calculate the pressure and density of polarized non-relativistic systems of 
two-component fermions coupled via a contact interaction at finite temperature.
For the unpolarized one-dimensional system with an attractive interaction,
we perform a third-order lattice perturbation theory calculation and assess its convergence by 
comparing with hybrid Monte Carlo. In that regime, we also demonstrate agreement with real Langevin. For the repulsive unpolarized one-dimensional system,
where there is a so-called complex phase problem, we present lattice perturbation theory as well as complex Langevin calculations. 
For our studies, we employ a Hubbard-Stratonovich transformation to decouple the 
interaction and automate the application of Wick's theorem for perturbative calculations, which generates the diagrammatic expansion at any order. 
We find excellent agreement between the results from our perturbative calculations and stochastic studies in
the weakly interacting regime. In addition, we show predictions for the strong coupling regime as well as for the polarized one-dimensional system. 
Finally, we show a first estimate for the 
equation of state in three dimensions where we focus on the polarized unitary Fermi gas.
}
\maketitle
\section{Introduction}\label{intro}
It is well-known that a large number of physically interesting quantum many-body systems are not amenable to being studied with stochastic techniques due to the appearance of the sign problem. For instance, relativistic systems at finite quark chemical potential and non-relativistic ones with a chemical potential asymmetry cannot use conventional quantum Monte Carlo techniques. A variety of techniques to circumvent, or at least mitigate, the sign problem in particular cases have been proposed and studied over the past several decades (see, for instance, Ref. \cite{GattringerSignReview} for a review). In this proceeding, we address advances on both perturbative and non-perturbative methods to compute the equation of state (EOS) for a many-body system of spin-1/2 particles under two-body contact interactions in situations normally hampered by a 
sign (or even complex-phase) problem. These systems can be physically realized as ultracold atoms, which provide a clean and malleable experimental situation to benchmark methods as well as many-body theories (see, e.g., \cite{Review1,Review2}).

Apart from Sec.~\ref{sec:3d}, where we consider the case of a three-dimensional polarized system in the so-called unitary limit, 
we study a polarized non-relativistic gas in one spatial dimension (1D), whose Hamiltonian is given by the 
Gaudin-Yang model~\cite{GAUDIN196755, PhysRevLett.19.1312}:
\begin{equation}
\label{Eq:Hamiltonian}
\hat{H} = -\frac{\hbar^2}{2m} \sum_i \frac{\partial^2}{\partial x_i^2} - g \sum_{i < j} \delta(x_i - x_j)\,,
\end{equation}
where $g$ is the bare coupling (directly related to the $s$-wave scattering $a_s$ length by $g = 2/a_s$) and we use units where $\hbar = m = 1$. 
In our perturbative and stochastic formalisms, the partition function $\mathcal{Z}$ is written as a path integral over a Hubbard-Stratonovich auxiliary field $\sigma$, such that
\begin{equation}
\label{Eq:HSTransform}
\mathcal{Z} = \int \mathcal{D}\sigma \, \det M_\uparrow(\sigma) \det M_\downarrow(\sigma)\,.
\end{equation}
The auxiliary field is placed on a lattice of extent $N_x^d \times N_\tau$, where $d$ is the spatial dimension and $\tau$ is the temporal lattice spacing; a 
Trotter-Suzuki decomposition has been performed such that $\beta = \tau N_\tau$, where $\beta$ is the inverse temperature.

In the following sections, we evaluate the above form of the path integral and display hybrid Monte Carlo (HMC), complex Langevin (CL), and lattice perturbation theory results for the pressure $P$ 
and particle density $n$ for the interacting Fermi gas in 1D at finite temperature. 
Details of the lattice perturbation theory and CL formalisms, as well as further comments on the results for the unpolarized system, can be found in Ref. \cite{PhysRevD.95.094502}. 
Additionally, we provide a first
estimate for the finite-temperature particle density of the polarized Fermi gas at unitarity 
in Sec.~\ref{sec:3d}, which corresponds to Eq. (\ref{Eq:Hamiltonian}) extended to three spatial dimensions and tuned to the threshold of bound-state formation.

\section{Complex Langevin formalism}\label{sec-1}

Conventional quantum Monte Carlo techniques, such as hybrid Monte Carlo (HMC) (see e.g. Ref.~\cite{QMCReview1}), 
typically rely on using a positive-definite probability measure $P(\sigma)$ in order to propose new field configurations via the Metropolis 
algorithm without a sign problem. In CL (see, e.g., Ref.~\cite{CL1}), the auxiliary field 
$\sigma$ is taken to be complex, such that $\sigma \rightarrow \sigma_{R} + i \sigma_{I}$. The dynamical equations of 
motion for the real and imaginary components of the field, in contrast to those that appear in HMC, are
\begin{eqnarray}
\delta \sigma_R &=& -\mathrm{Re}\left[\frac{\delta S(\sigma)}{\delta \sigma}\right]\delta t + \eta \sqrt{\delta t}\,,\\
\delta \sigma_I &=& -\mathrm{Im}\left[\frac{\delta S(\sigma)}{\delta \sigma}\right]\delta t\,,
\end{eqnarray}
where we define the action $S = -\ln ({\det} M_{\uparrow}[\sigma] \, {\det}  M_{\downarrow}[\sigma])$ and $\eta$ is 
a $t$-dependent noise term. 
{Loosely speaking, the equilibrium distribution of the auxiliary field obtained from the CL equations is then assumed to be identical 
to the probability measure associated with the path integral~\eqref{Eq:HSTransform} and can therefore be exploited to compute physical
observables. Note that, in the case where the imaginary part of $\sigma$ vanishes, CL reduces to real Langevin (RL).
Unlike HMC (and also RL), however, it is} understood that CL generally may not be guaranteed to converge to the correct result, if at all, unless certain 
criteria are satisfied~\cite{CL2,Aarts:2017vrv}; as such, additional scrutiny is required to ensure correct results. One of the most common challenges in 
CL calculations are uncontrolled excursions of the auxiliary field into the complex plane due to singularities that appear in 
the determinants of~$M_{\uparrow,\downarrow}$. For the system 
we study here, we have found that it is indeed necessary to modify the equations of 
motion of the auxiliary field to avoid such excursions. We do so by adding a regulating term controlled by the real parameter $\xi$, such that we obtain
\begin{eqnarray}
\delta\sigma_R &\rightarrow& \delta\sigma_R - 2 \xi \sigma_R \delta t\,,\\
\delta\sigma_I &\rightarrow& \delta\sigma_I - 2\xi \sigma _I \delta t\,.
\end{eqnarray}
In the case where $\xi = 0$ and the regulating term vanishes, the calculation becomes uncontrolled and the normalized density appears to diverge. 
However, for a moderately-sized window about $\xi \simeq 0.1$, the calculation becomes well-controlled and converges quickly to the expected value,
see also Ref.~\cite{Rammelmuller:2017vqn} for a detailed study of the role of this parameter for mass-balanced Fermi gases.
 The distribution of values of the determinant also indicates that zeros in the determinants are not seen, at least for the couplings
studied.
For strong repulsive couplings, however, our studies of this regime in the ground state indicate that the probability distributions associated with physical observables may not exhibit a finite variance, regardless of the polarization, which suggests the appearance of an overlap problem. Detailed studies are required in the future to gain a comprehensive understanding of this issue.

\section{Perturbation theory formalism}

\begin{figure}[h]
    \centering
    \includegraphics[width=14cm]{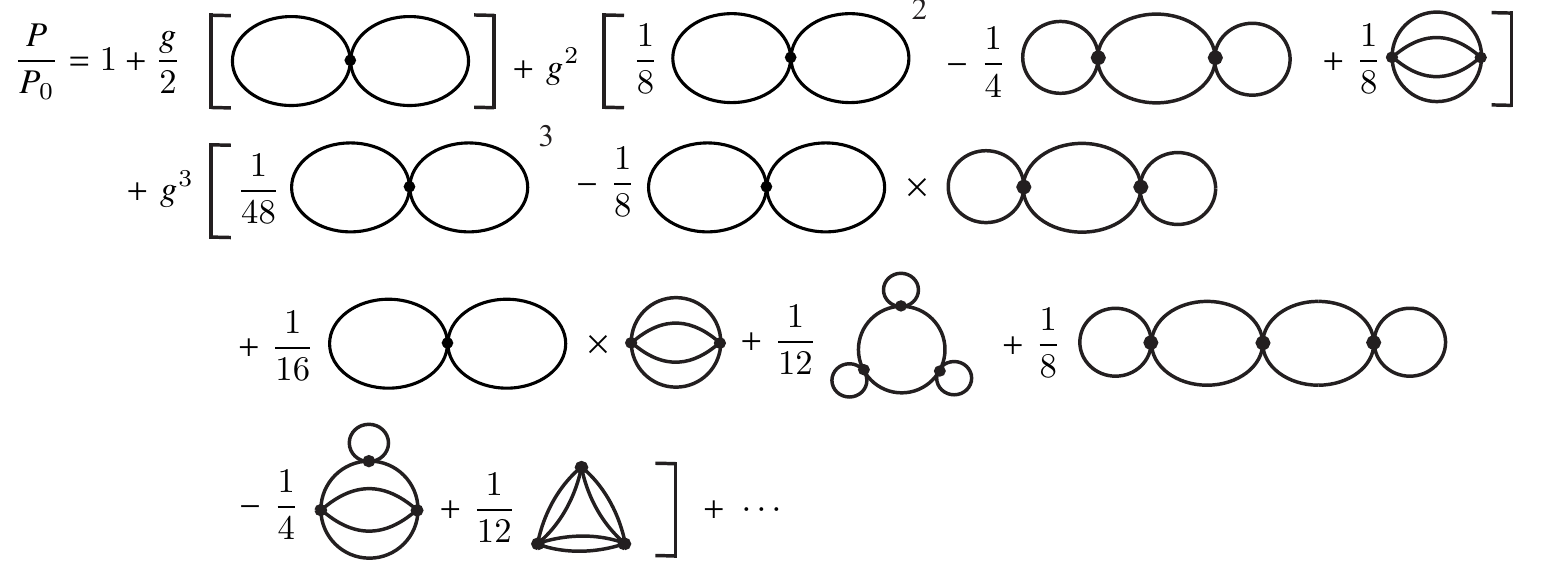}
    \caption{Expression for the perturbative expansion of the normalized pressure $P/P_0$ in terms of first-order (NLO), second-order (N2LO), and third-order (N3LO) diagrams and associated symmetry factors. As written, the pressure is expanded in the coupling $g$ (see Ref. \cite{PhysRevD.95.094502} for details).}
    \label{Fig:Diagrams}
\end{figure}

In addition to the non-perturbative CL technique, we also developed a method for 
(semi-)analytically computing the perturbative expansion of the pressure which we applied 
to next-to-next-to-next-to-leading order~(N3LO). This expansion is performed on the lattice starting from the same 
Hubbard-Stratonovich form of the partition function as given
in Eq. (\ref{Eq:HSTransform}). Since both methods are computed on the lattice, results can be directly compared for a given system volume. 
For illustration purposes, let us consider the unpolarized system.
There, the fermion determinant $M$ is expanded in powers of the coupling about the non-interacting limit as follows:
\begin{equation}
\mathcal{Z} = \int \mathcal{D}\sigma \, {\det}^2 M_0 \, [1 + g \, f_1(\sigma) + g^2 f_2(\sigma) + \cdots]^2\,.
\end{equation}
Here, $M_0$ is 
the fermion determinant corresponding to the non-interacting system ($g = 0$), and $f_i(\sigma)$ are determined functions over which the path integral is computed exactly. The functions $f_i(\sigma)$ correspond to a sum of $i$-th order Feynman diagrams with associated symmetry factors (see Ref. \cite{PhysRevD.95.094502} for details).
The resulting perturbative expansion 
for the pressure $P$ up to N3LO normalized by the non-interacting counterpart 
$P_0$ in terms of Feynman diagrams is displayed in Fig. \ref{Fig:Diagrams}.

\section{Results for systems in one dimension}
In the following section, we will display results for the density of both unpolarized and spin-polarized systems in one spatial dimension. 
Computations using both the perturbative and stochastic techniques will be shown.

\subsection{Unpolarized system}
\begin{figure}[t]
  \centering
  \includegraphics[width=7cm,clip]{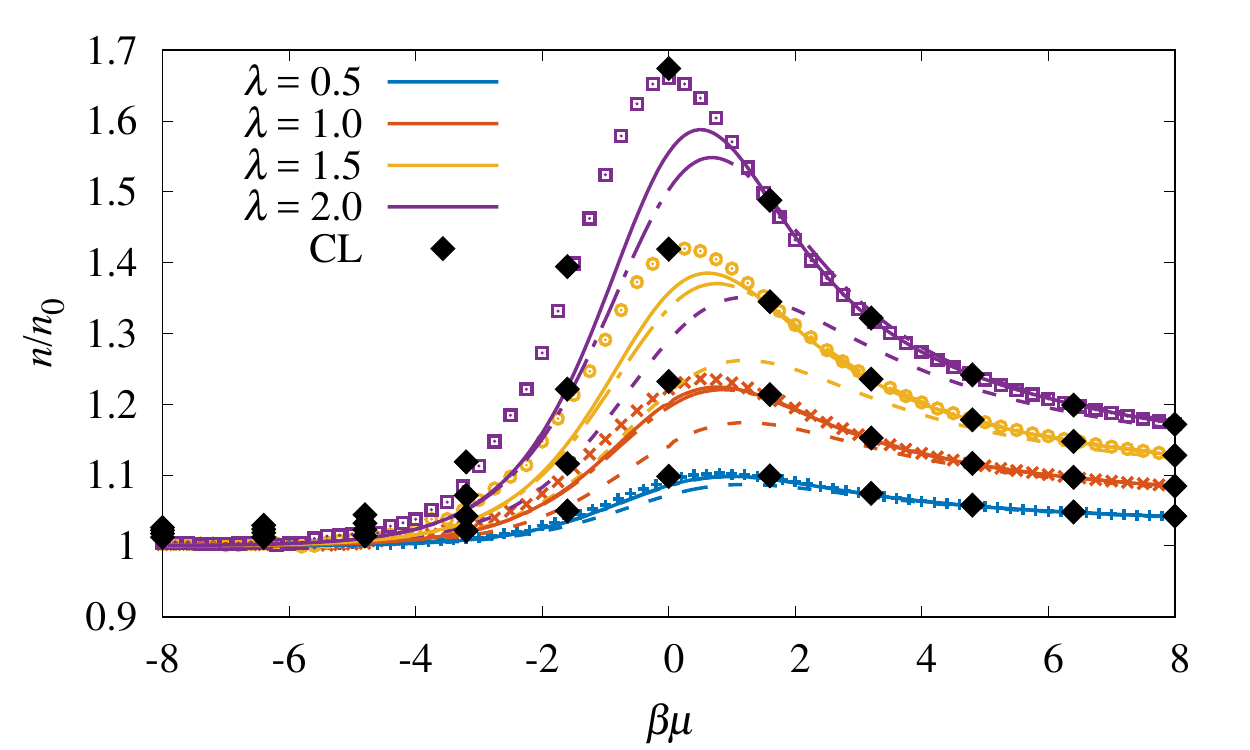}
    \includegraphics[width=7cm,clip]{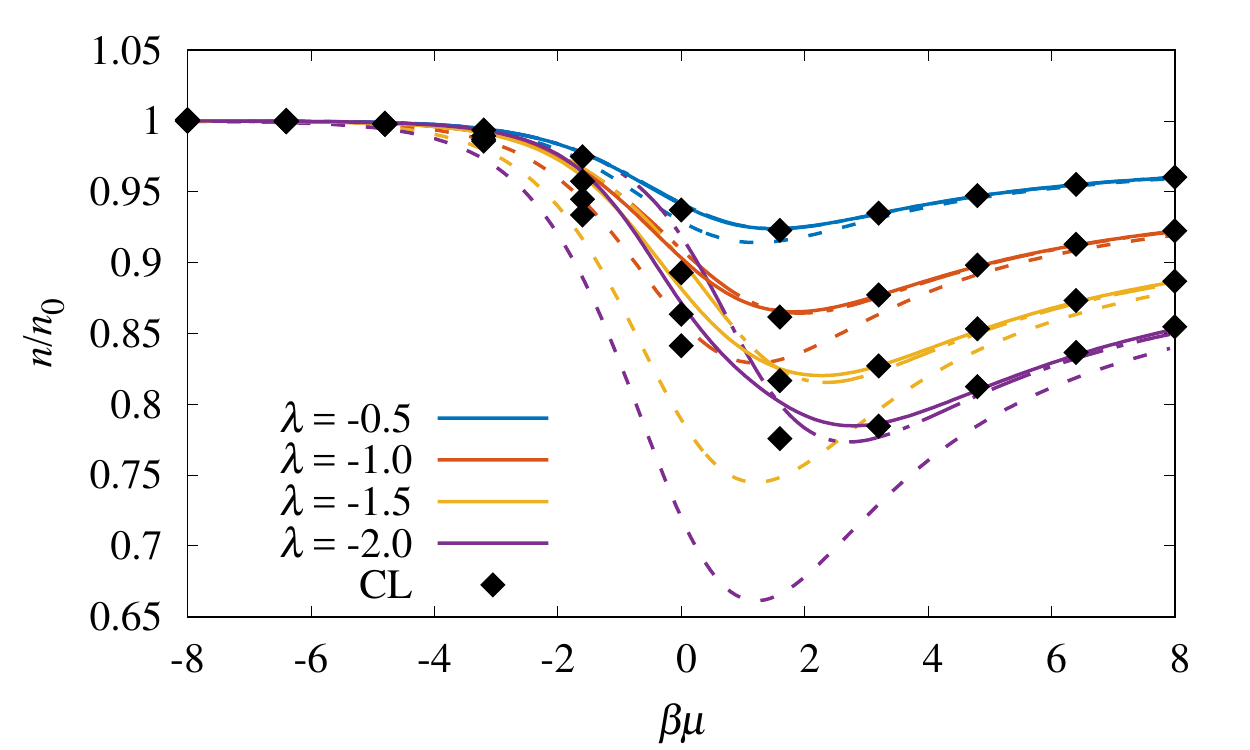}
  \caption{Density $n$ of the attractive (left) and repulsive (right) unpolarized Fermi gas normalized by the density of the noninteracting system $n_0$. Results are shown for the dimensionless attractive and repulsive interaction strengths $\lambda=\sqrt{\beta}g = \pm 0.5, \pm 1.0, \pm 1.5, \pm 2.0$. The NLO (dashed line), N2LO (dash-dotted line), and N3LO (solid line) results are displayed for each coupling and are 
  compared with HMC results {(depicted by symbols except black diamonds)} for the attractive case {($\lambda > 0)$, see Ref. \cite{PhysRevA.91.033618}}.
  For both plots, the black diamonds show CL (RL for the attractive case), regulated with $\xi = 0.1$, as described in the main text. 
  Note also that the N2LO and the N3LO results already agree over a wide range of values for~$\beta\mu$ for the considered values 
  of the coupling.}
  \label{Fig:UnpolarizedDensity}
\end{figure}

\begin{figure}[t] 
  \centering
  \includegraphics[width=7cm,clip]{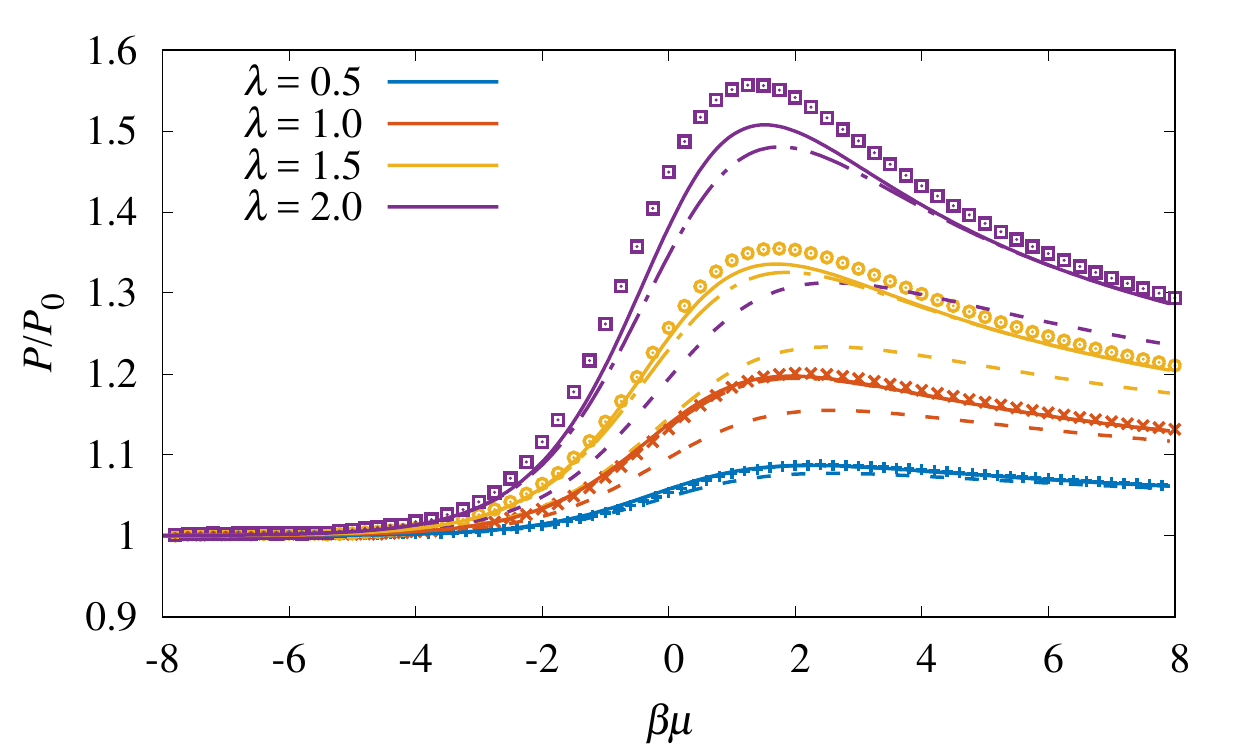}
    \includegraphics[width=7cm,clip]{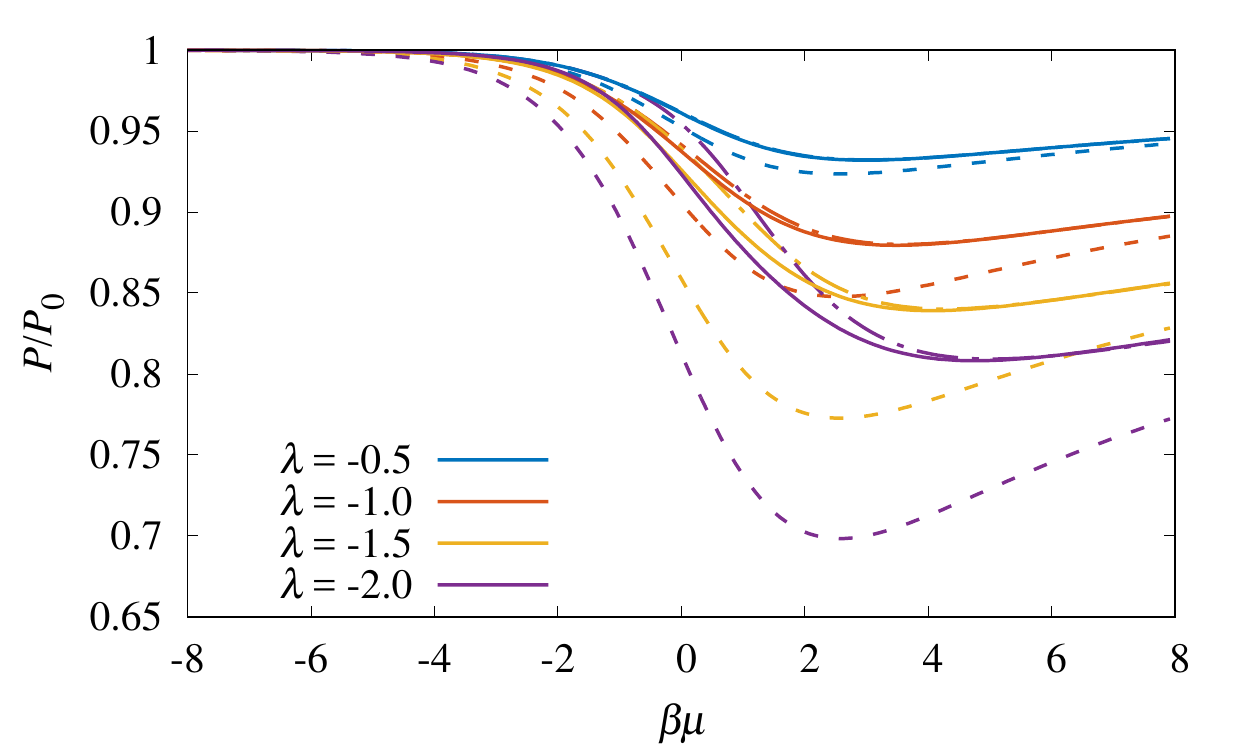}
  \caption{Perturbation theory results for the pressure $P$ of the attractive (left) and repulsive (right) unpolarized Fermi gas normalized by the pressure of the non-interacting system $P_0$. Results are shown for the dimensionless attractive and repulsive interaction strengths 
  {$\lambda =\sqrt{\beta}g = \pm 0.5, \pm 1.0, \pm 1.5, \pm 2.0$. The} 
  NLO (dashed line), N2LO (dash-dotted line), and N3LO (solid line) are displayed for each coupling. 
  The corresponding data points for each attractive coupling (depicted by symbols) are computed using HMC (see Ref. \cite{PhysRevA.91.033618}).
  Note again that the N2LO and the N3LO results already agree over a wide range of values for~$\beta\mu$ for the considered values 
  of the coupling.}
  \label{Fig:UnpolarizedPressure}
\end{figure}

In the case of the attractive unpolarized Fermi gas as described by identical chemical potentials of the two species, 
$\mu_\uparrow = \mu_\downarrow$, a sign problem is not present in conventional HMC 
calculations.
In the repulsive case, however, $g < 0$ [see our convention for the sign of the coupling in Eq.~\eqref{Eq:Hamiltonian}],
and a sign problem is present since the matrices $M_{\uparrow,\downarrow}$ are generally complex. 
It should be pointed out that other methods exist that can address the sign problem in this unpolarized case, but they are
restricted to 1D, whereas
our present work aims at the development of a toolbox that can be applied in any dimension.
Fig.~\ref{Fig:UnpolarizedDensity} displays the particle number density $n$ normalized by its noninteracting counterpart $n_0$ for the attractive and 
repulsive cases, and Fig.~\ref{Fig:UnpolarizedPressure} displays the corresponding pressure $P$, normalized by the noninteracting counterpart $P_0$.

As can be appreciated in both figures, the agreement in Fig.~\ref{Fig:UnpolarizedDensity} (left panel) between HMC and RL for the attractive 
case is excellent, as expected from a situation that does not feature a sign problem. This agreement is encouraging as a test of our understanding of 
Langevin-based methods. The agreement with perturbative results is also remarkable, except in the deep quantum region around $\beta\mu = 0$ and at 
strong coupling, where some deviations are expected and indeed found. Proceeding to 
the repulsive case (right panel of Fig.~\ref{Fig:UnpolarizedDensity}), where HMC is not applicable, we see that CL results show once again excellent agreement with perturbation theory at weak coupling, and as the order in perturbation theory is increased, the agreement with CL improves substantially. We regard this situation as strongly supporting the use of CL for repulsively interacting, unpolarized 
gases, at least for the considered range of coupling strengths. 
To the best of our knowledge, the finite-temperature EOS of the repulsive Fermi gas in 1D presented in 
Fig.~\ref{Fig:UnpolarizedDensity} (right panel) (which appears in our work of Ref.~\cite{PhysRevD.95.094502}) is the first time such a calculation was carried out. From the density, by integration over $\beta \mu$, we obtain the pressure, which is shown in Fig.~\ref{Fig:UnpolarizedPressure}. While we do not have at the moment enough precision in our CL data to carry out the integration for the repulsive case, we use this opportunity to attempt a perturbative calculation and compare with HMC on the attractive side. The agreement in the latter case is excellent, especially at weak coupling (as expected), which encourages us to conclude that (also based on our results for the density) the pressure on the repulsive side is well captured by perturbation theory for the couplings we explored.

%

\subsection{Polarized system} 
\begin{figure}[t]
  \centering
  \includegraphics[width=7cm,clip]{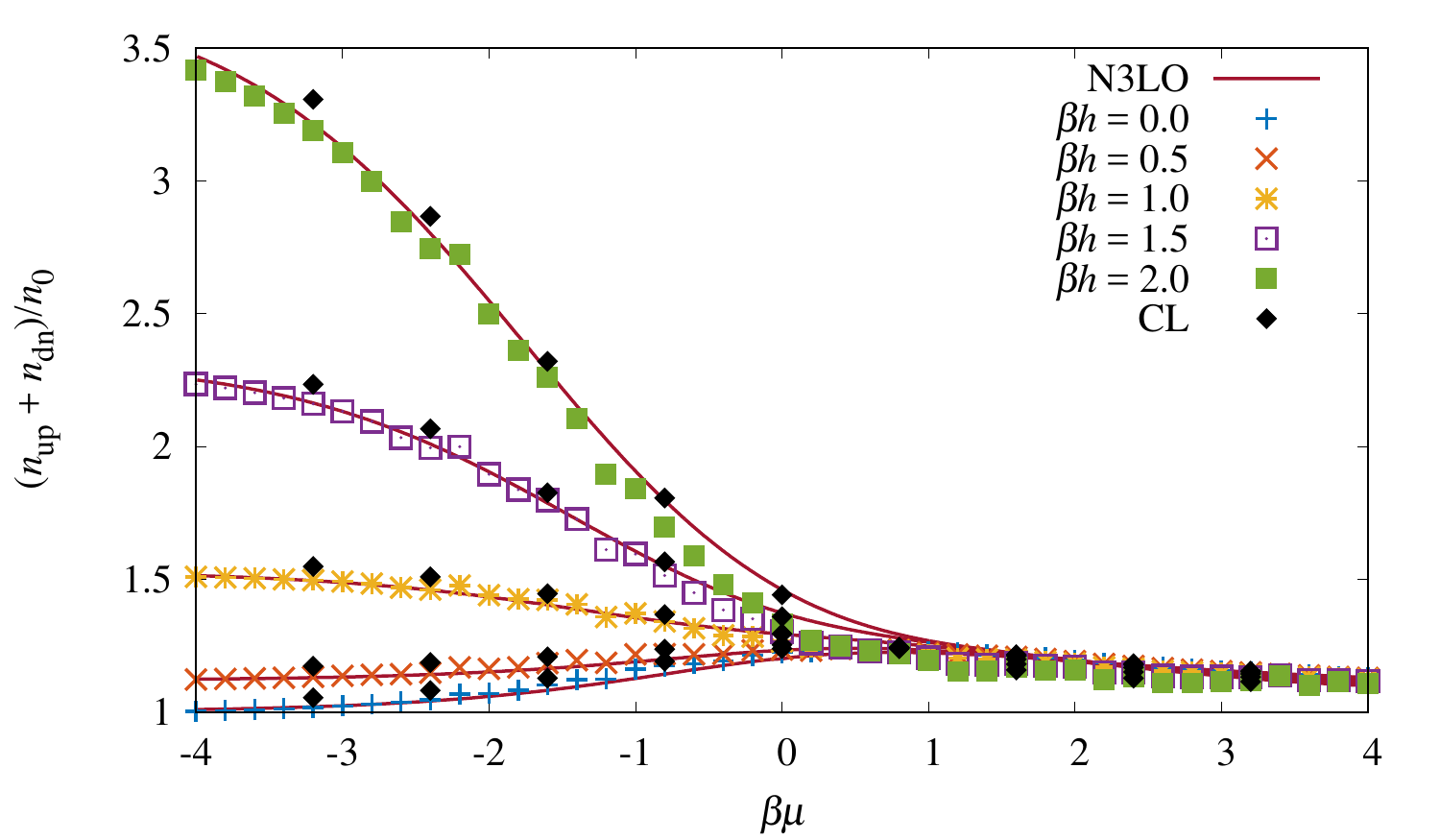}
    \includegraphics[width=7cm,clip]{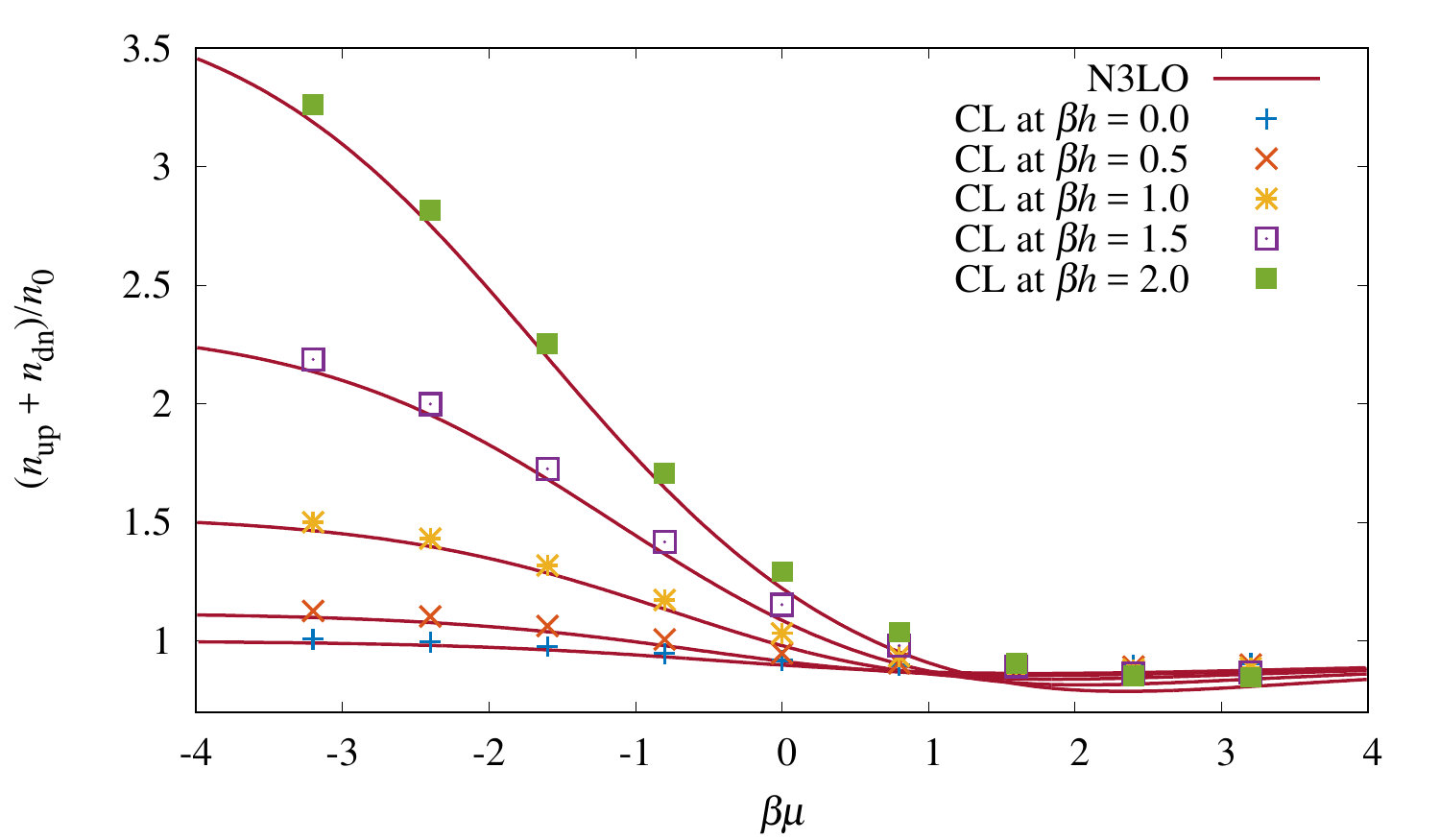}
  \caption{Density $n$ of the attractive (left) and repulsive (right) spin-polarized Fermi gas normalized by the density of the noninteracting, 
  unpolarized system $n_0$. Results are shown for the dimensionless attractive and repulsive interaction strength $\lambda = \pm 1$ and 
  for chemical potential asymmetries $\beta h = 0.0, 0.5, 1.0, 1.5, 2.0$. 
  {In both panels, the red solid lines correspond to perturbation theory results at N3LO for the various values of~$\beta h$. In the left panel, the} 
  solid black {diamonds correspond to CL results}, 
  and the colored symbols correspond to to HMC {results obtained by using imaginary polarization (see Ref. \cite{PhysRevA.92.063609}). In the right panel depicting 
  the repulsive case, where HMC is not applicable, the colored symbols correspond to CL results.}}
  \label{Fig:PolarizedDensity}
\end{figure}

\begin{figure}[t]
	\centering
	\includegraphics[width=7cm,clip]{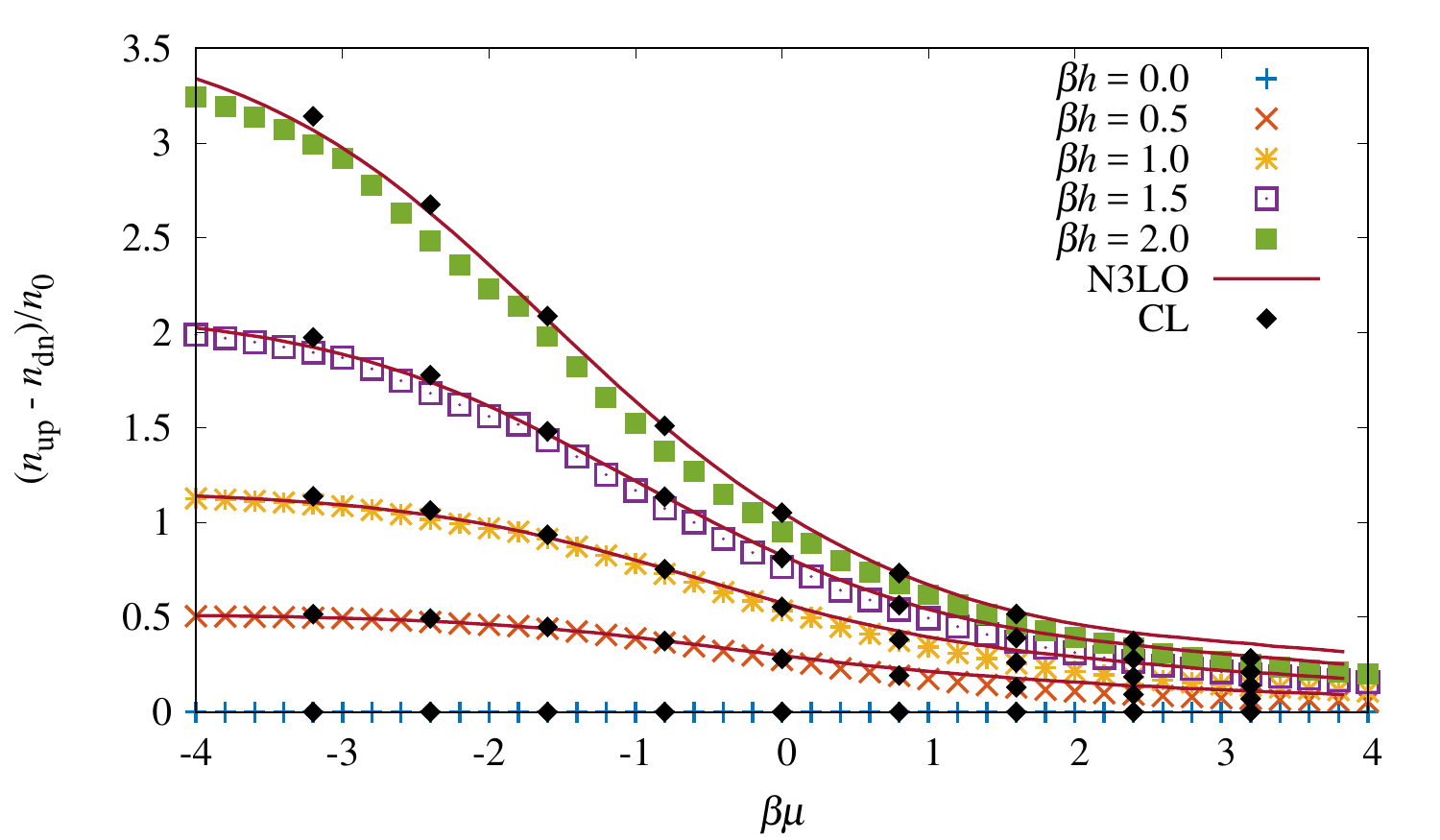}
	\includegraphics[width=7cm,clip]{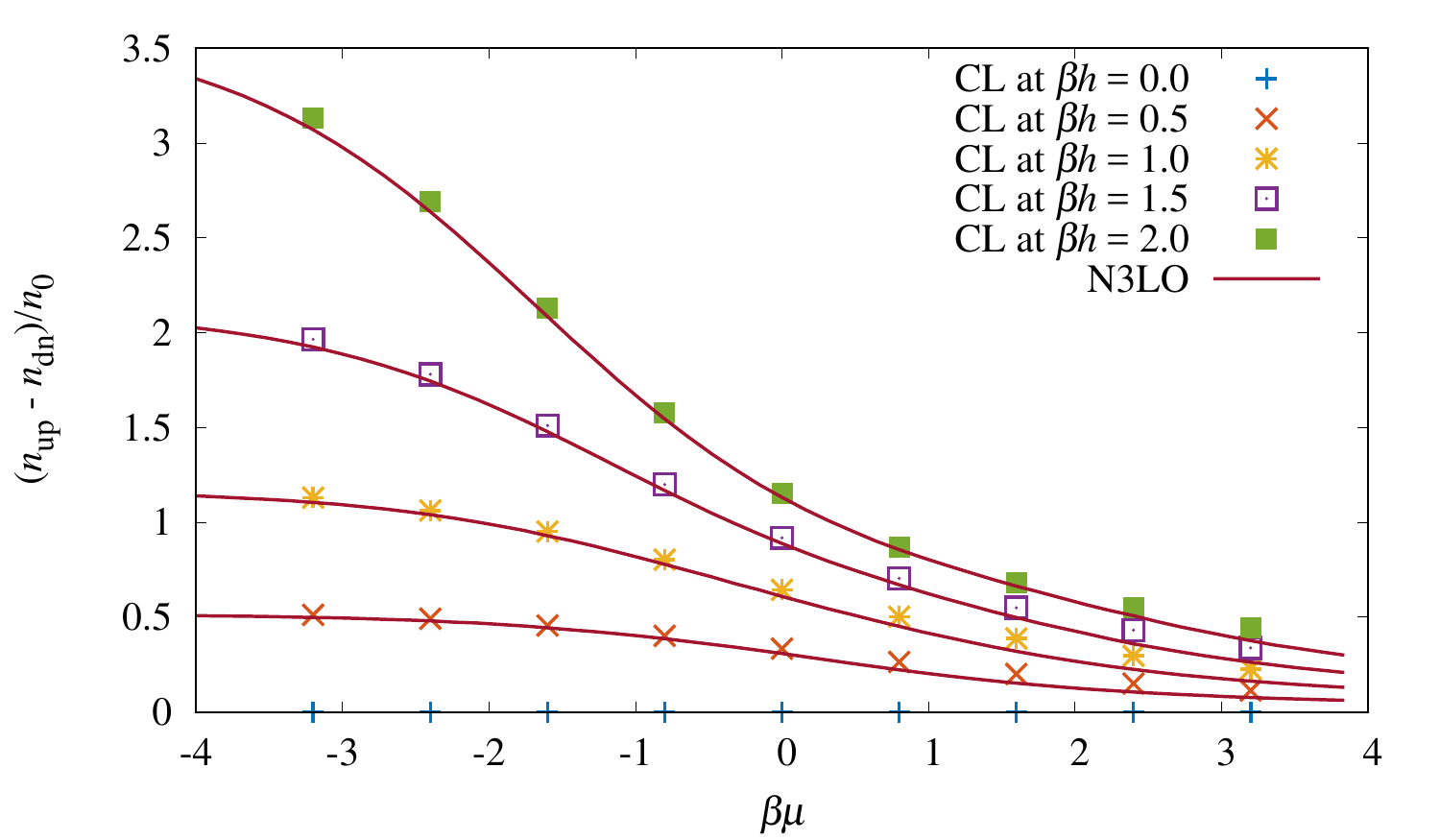}
	\caption{Magnetization $m$ of the attractive (left) and repulsive (right) spin-polarized Fermi gas normalized by the density of the noninteracting, unpolarized system $n_0$. Results are shown for the dimensionless attractive and repulsive interaction strength $\lambda = \pm 1$ and for chemical potential asymmetries $\beta h = 0.0, 0.5, 1.0, 1.5, 2.0$. {In both panels, the red solid 
	lines correspond to perturbation theory results at N3LO for the various values of~$\beta h$. In the left panel, the} 
  solid black {diamonds correspond to CL results, 
  and the colored symbols correspond to HMC results obtained by} using imaginary polarization (see Ref. \cite{PhysRevA.92.063609}). 
  {In the right panel depicting
  the repulsive case, where HMC is not applicable, the colored symbols correspond to CL results.}}
	\label{Fig:PolarizedMagnetization}
\end{figure}

The spin-polarized Fermi gas, where $\mu_\uparrow \neq \mu_\downarrow$, has a sign problem for both the 
attractive and repulsive cases, and it cannot be completely avoided with any other 
stochastic method that we know of, in any dimension. 
One way to proceed is to apply HMC to a system where we introduce a complex chemical potential 
such that the asymmetry $h = (\mu_\uparrow - \mu_\downarrow)/2$ is imaginary, after which the observables are obtained via analytic continuation. This method was put forward for non-relativistic systems in Ref.~\cite{PolarizedUFG3}
and first applied to the system studied here in Ref.~\cite{PhysRevA.92.063609}. 
For a variety of dimensionless chemical potential asymmetries~$\beta h$, the normalized total density $n/n_0 = (n_\uparrow + n_\downarrow)/n_0$
obtained from this approach is displayed in Fig.~\ref{Fig:PolarizedDensity} for both the attractive and repulsive cases. Similarly, the magnetization 
$m/n_0 = (n_\uparrow - n_\downarrow)/n_0$ is shown in Fig.~\ref{Fig:PolarizedMagnetization}.

In both panels of Fig.~\ref{Fig:PolarizedDensity}, the agreement among the methods is qualitatively and 
quantitatively satisfactory. The scale at large negative $\beta\mu$ is governed by the virial expansion, which predicts 
the same results $n/n_0 = \cosh (\beta h)$ at sufficiently small $z={\rm e}^{\beta \mu}$, regardless of the form of the interaction; the agreement with that result is very good.
Finally, Fig.~\ref{Fig:PolarizedMagnetization} shows a similar situation for the magnetization, whose behavior at small $z$ is $m/n_0 = \sinh (\beta h)$.
For both attractive and repulsive couplings, however, the differences in the results from the various methods increase in some regimes.
This does not come unexpected in the regime of large imbalances. Recall that 
an analytic continuation of the numerical data points underlies the HMC computations of physical observables in this case.
Such an analytic continuation appears to become increasingly unreliable for large imbalances; see Ref.~\cite{PhysRevA.92.063609} 
and also Refs.~\cite{Braun:2014pka,Rammelmuller:2017vqn} where this has also been
observed in the case of imaginary mass imbalances. Apart from that, an understanding of the emergence of the discrepancy 
between CL and perturbation theory for repulsive interactions and $\beta \mu > 0$ with increasing imbalances requires a more detailed analysis, also with respect to the
applicability of CL in this regime~\cite{CL2,Aarts:2017vrv}.
%

\section{First glance at the EOS of the unitary Fermi gas\label{sec:3d}}

\begin{wrapfigure}{hr}{70mm}
	\vspace{-20pt}
	\begin{center}
		\includegraphics[width=0.5\columnwidth]{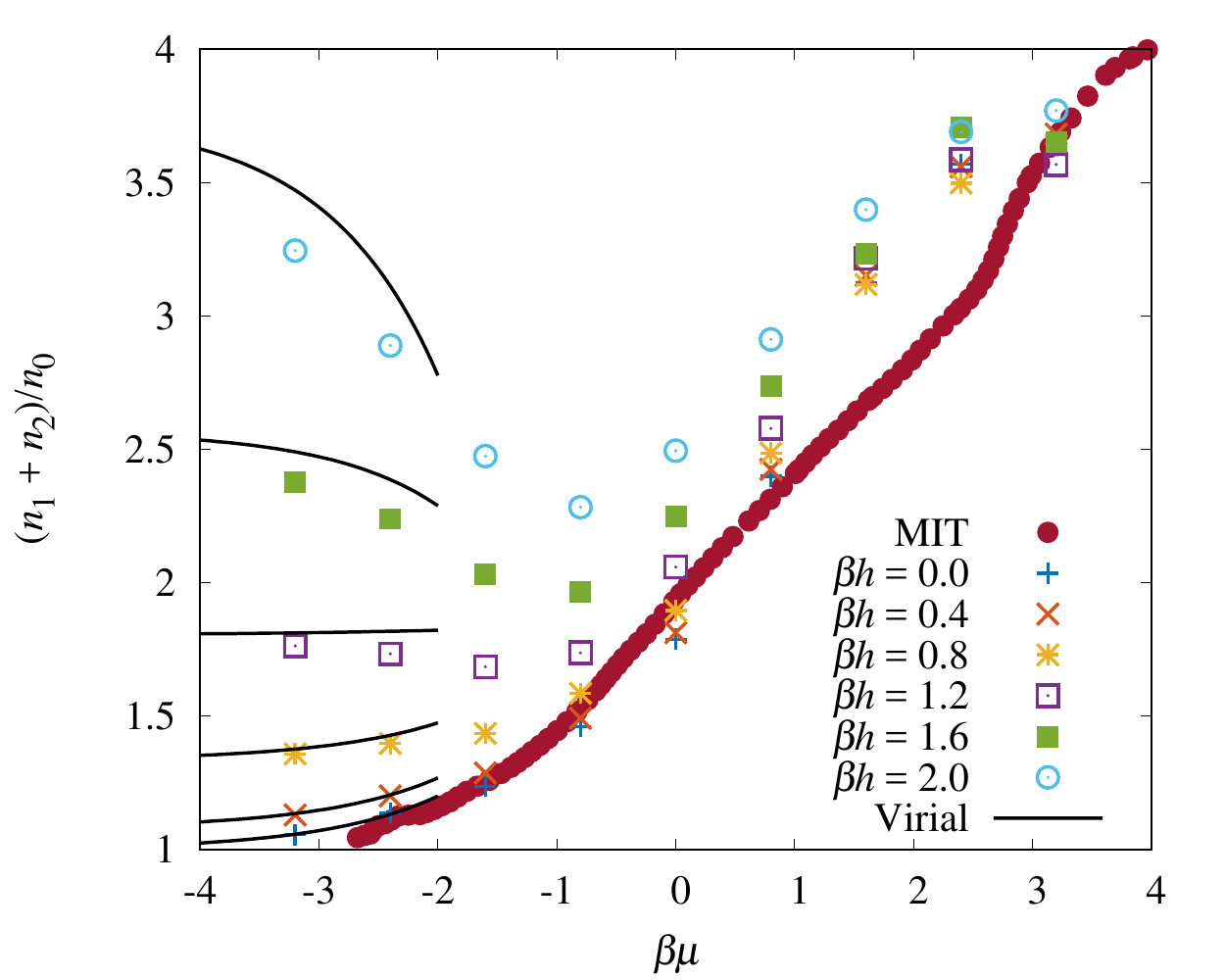}
	\end{center}
	\vspace{-15pt}
\caption{\label{Fig:PolarizedUFG} 
{Density $n$ of the} attractively-interacting, spin-polarized Fermi gas at unitarity in three spatial dimensions, normalized by the density of the noninteracting, unpolarized system $n_0$. Results using CL are shown in colored points for chemical potential asymmetries $\beta h = 0.0, 0.4, 0.8, 1.2, 1.6, 2.0$ at a spatial lattice volume of $N_x^3 = 7^3$. The second-order virial expansion is shown in solid black lines. Additionally, experimental results for the unpolarized system at unitarity are shown in solid red circles (see Ref. \cite{MITUFG}).
}
\vspace{-15pt}
\end{wrapfigure}
Encouraged by our results in 1D, we proceed to a special 
3D case known as the unitary Fermi gas (UFG). This system is a two-component gas with a contact interaction tuned to the threshold of bound-state formation. 
At that point, the $s$-wave scattering length, which controls the strength of the interaction, 
is infinite, such that the many-body problem features as many scales as a non-interacting gas, yet its behavior is that of a strongly correlated system.
While much is known about the UFG (see, e.g., Refs.~\cite{ZwergerBook, BDM1, BDM2, BDM3}), especially about its 
{equation of state in the unpolarized case, its} polarized state
remains a topic of active 
{investigation, see, e.g.,} Refs.~\cite{PolarizedUFG1,PolarizedUFG2,PolarizedUFG3,PolarizedUFG4,PolarizedUFG5,PolarizedUFG6}. 
In Fig.~\ref{Fig:PolarizedUFG}, we show preliminary CL results for the density of the UFG as a function of 
$\beta\mu$ for various asymmetries $\beta h$. The same figure shows a comparison with experiment (unpolarized) (see Ref.~\cite{MITUFG}) as well as with the second-order virial expansion. While the agreement with the latter is encouraging, some sizable deviations from experiment remain at $\beta \mu > 0$ {in the unpolarized case}, 
possibly due to finite-volume effects, which are currently under investigation. 
Nevertheless, the overall qualitative agreement is encouraging. Once those discrepancies are resolved, we plan to explore the phase diagram of polarized superfluid matter, which has been conjectured (but not yet solidly proven) to contain exotic phases.

\section{Conclusion}

In this contribution we address the basic equations of state (density and polarization) of strongly {coupled non-relativistic matter. 
Repulsion} and polarization, each by itself, would typically yield a sign problem 
for auxiliary-field {methods as considered in the present studies for 1D systems, and for} all methods we know in higher dimensions. For that reason, we resorted to the CL method, which 
shows excellent agreement with third-order perturbative results. Encouraged by those results, we set out to explore with CL the  
{EOS} of a strongly interacting 3D system at finite polarization, namely the UFG. The results for the latter show promise in that they  
agree at least qualitatively with the virial expansion and {with experimental data for the unpolarized system. Studies} of systematic effects are underway.

{
The authors acknowledge useful discussions with Lukas Rammelm{\"u}ller
and collaboration on related studies.
This material is based upon work supported by the
National Science Foundation under Grants No.
PHY{1452635} (Computational Physics Program) and
DGE{1144081} (Graduate Research Fellowship Program). J.B. acknowledges 
support by HIC for FAIR within the LOEWE program of the State of Hesse. 
Numerical calculations have partially been performed at the LOEWE-CSC Frankfurt.
}


\clearpage
\bibliography{lattice2017}

\end{document}